\documentclass[conference]{IEEEtran}
\IEEEoverridecommandlockouts

\usepackage[left=0.6in, right=0.6in, top=0.7in, bottom=0.6in]{geometry} 
\usepackage{cite}
\usepackage{amsmath,amssymb,amsfonts}
\usepackage{algorithmic}
\usepackage{algorithm}
\usepackage{graphicx}
\usepackage{array}
\usepackage{textcomp}
\usepackage{xcolor}
\def\BibTeX{{\rm B\kern-.05em{\sc i\kern-.025em b}\kern-.08em
    T\kern-.1667em\lower.7ex\hbox{E}\kern-.125emX}}

\begin{document}

\title{Dynamic Task and Resource Scheduling Towards Green Space-Air-Ground-Sea Integrated Network\\

}

\author{
\IEEEauthorblockN{Yufei Ye\textsuperscript{1}, Shijian Gao\textsuperscript{2}, Xinhu Zheng\textsuperscript{1,2}, and Liuqing Yang\textsuperscript{1,2}}
\IEEEauthorblockA{\textsuperscript{1}Intelligent Transportation Thrust, Hong Kong University of Science and Technology (Guangzhou), Guangzhou, China}
\IEEEauthorblockA{\textsuperscript{2}Internet of Things Thrust, Hong Kong University of Science and Technology (Guangzhou), Guangzhou, China}
\IEEEauthorblockA{Email: yye760@connect.hkust-gz.edu.cn, \{shijiangao, xinhuzheng, lqyang\}@hkust-gz.edu.cn}
}

\maketitle

\begin{abstract}
In the context of 6G ubiquitous connectivity, the space–air–ground–sea integrated network (SAGSIN) emerges as a new paradigm to provide critical services for resource-limited ocean environments. To realize this paradigm efficiently, we propose an innovative dynamic task and resource scheduling approach for green SAGSIN that delivers computing support for vessels while minimizing overall task execution delay. To address the challenge of multi-layer task scheduling, a layer-wise task offloading algorithm is developed specifically for SAGSIN. It adapts to real-time, multi-dimensional system dynamics and integrates an anticipatory handover strategy that adaptively controls the amount of data offloaded to the satellite, thereby preventing post-handover congestion while improving satellite resource utilization. Furthermore, the bandwidth allocation of uncrewed aerial vehicles and base station, UAV trajectories, and computing resource allocation are jointly optimized to enhance connectivity among low‑altitude devices and facilitate demand‑driven resource allocation for green network development. Simulation results verify that the proposed method better adapts to dynamic system resources and achieves at least a 23\% reduction in average task delay compared with benchmarks.

\end{abstract}
\begin{IEEEkeywords}
Space-air-ground-sea integrated network, task offloading, efficient scheduling, edge computing, satellite handover.
\end{IEEEkeywords}
\vspace{-0.35cm}

\section{Introduction}

Driven by the vision of 6G ubiquitous connectivity, expanding maritime activities, such as environmental monitoring and resource exploration, create growing demands for communication and computing resources \cite{jyou2025joint}. However, the limited computing power of vessels and the scarce maritime infrastructure hinder the support for maritime applications. Leveraging the flexible mobility and deployment, uncrewed aerial vehicles (UAVs), which are key enablers of the emerging low‑altitude economy (LAE), have attracted attention as a means of providing computing support to vessels \cite{ybai2025dynamic} and vehicles \cite{yye2025multitier}. However, due to the compact design, UAVs have limited computing capabilities and struggle to support intensive computational tasks. In this context, the space‑air‑ground‑sea integrated network (SAGSIN) has emerged as a promising paradigm for maritime edge computing \cite{sgao2026integrated}. It leverages the complementary advantages of heterogeneous networks and integrates their resources to provide computing services for vessels, thereby alleviating the energy and computational burden of onboard devices.

When UAVs are assisted by coastal base stations (BSs), maritime task execution delay can be minimized by optimizing task offloading ratios \cite{mdai2023latency, mli2026online} and caching decisions \cite{yzhang2025joint}. Leveraging the wide coverage and strong computing power of low earth orbit (LEO) satellites, UAV‑satellite collaborative computing frameworks have been designed to save energy for the entire system \cite{zwang2025double} or for UAVs alone \cite{sjung2023marine}. A scheme for tasks with different delay sensitivities was proposed in \cite{sqi2025joint}, and a satellite‑assisted harvesting‑and‑offloading method was developed to support UAVs in \cite{mdai2025energy}. Moreover, \cite{hzhang2025energy} designed a three‑tier computing system that augments satellites with base stations. Serving as relay nodes between the space and sea layers, high‑altitude platforms (HAPs) can effectively mitigate the high path loss of satellites. They were adopted in \cite{wli2026efficient} and \cite{wwu2025multi} to further reduce task costs. Meanwhile, task offloading methods for improving energy efficiency and communication security in SAGSIN were studied in \cite{zlin2024maritime} and \cite{dwang2023double}, respectively. Despite the potential of SAGSIN, most existing studies simplify the network architecture to cope with the high complexity of multi‑layer task offloading. This simplification may lead to inefficient resource utilization and load imbalance. Furthermore, fixed offloading and resource allocation strategies fail to adapt to rapid changes in system states during data transmission, such as fluctuations in available computing and bandwidth resources, which further degrades system performance. Additionally, the limited service time of fast‑moving satellites and the critical issue of inter‑satellite handover remain underexplored.

To overcome these limitations, this paper proposes a novel \underline{d}ynamic t\underline{a}sk and re\underline{s}ource sc\underline{h}eduling scheme (DASH) for green SAGSIN. Unlike static or heuristic offloading methods, DASH introduces a low‑complexity, layer‑wise dynamic task offloading algorithm based on backpressure routing theory. This algorithm adaptively adjusts fine‑grained task scheduling decisions according to real‑time system conditions, including available computing resources of heterogeneous servers, task congestion levels, link capacities, and network topology. To address the underexplored handover problem, DASH further incorporates an anticipatory strategy that proactively regulates satellite offloading traffic based on the joint states of both the current and incoming satellites, thereby preventing post‑handover congestion while improving satellite resource utilization. Beyond task offloading, we jointly optimize UAV‑BS bandwidth allocation, UAV trajectory planning, and computing resource allocation. This strengthens connectivity between vessels and low‑altitude UAVs and boosts task completion through the co‑design of resource scheduling and physical network architecture, a dimension largely overlooked in prior SAGSIN research. Through adaptive multi‑layer load balancing and demand‑driven resource allocation, DASH provides efficient computing services while significantly improving resource utilization efficiency and the ecological stability of the system, laying a crucial foundation for green and sustainable SAGSIN development. Simulation results demonstrate that DASH outperforms benchmarks in adapting to variations in computing and bandwidth resources, achieving at least a 23\% reduction in average task delay while effectively alleviating post‑handover satellite congestion.


\section{System Model and Problem Formulation}

In this section, we first outline the proposed SAGSIN system model, which encompasses the network modeling, communication modeling, and computation and task queuing modeling. Then we present the problem formulation and detail the overarching decomposition framework to facilitate problem-solving.

\begin{figure}[!t]
\centerline{\includegraphics[width=2.6 in]{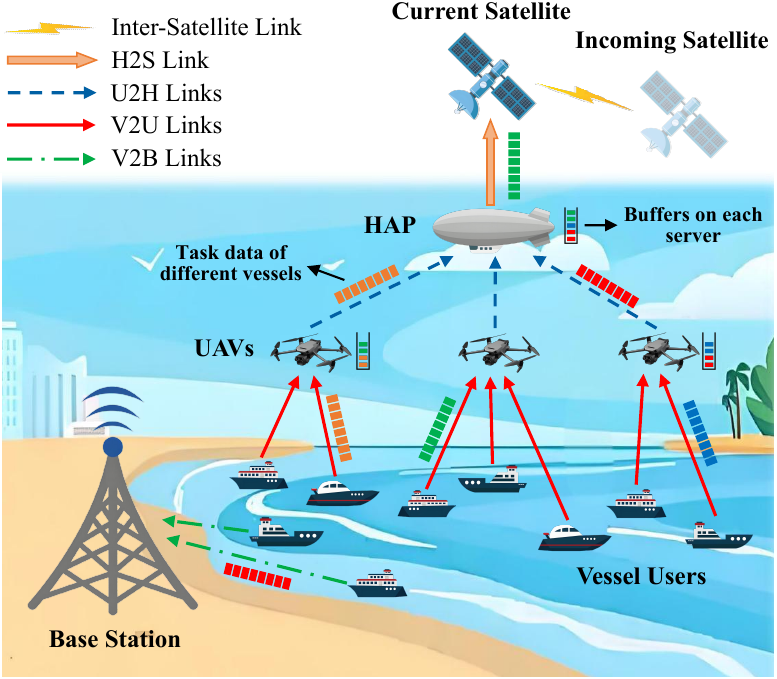}}
\vspace{-5pt}
\caption{Illustration of the multi-layer network architecture for SAGSIN.}
\label{system}
\end{figure}

\subsection{Network Model}

Fig.~\ref{system} depicts the proposed system, comprising $V$ vessels and $U$ UAVs, denoted by set $\mathcal{V}$ and $\mathcal{U}$, respectively, along with a coastal BS, an HAP, a current and an incoming LEO satellite. We discretize time into slots, each indexed by $n$ with length of $\tau$. The horizontal locations of vessel $v$, UAV $u$, BS, and HAP are denoted by $\boldsymbol{W_v}(n)$, $\boldsymbol{W_u}(n)$, $\boldsymbol{W_b}$, and $\boldsymbol{W_h}(n)$, with the height of $0$, $H_u$, $H_b$, and $H_h$, respectively. The displacement of UAV $u$ between adjacent time slots is $\Delta d_{u}(n)=\|\boldsymbol{W_u}(n)-\boldsymbol{W_u}(n-1)\|$. The size and computational density of vessel $v$'s task data are $D_v$ bits and $C_v$ cycles/bit, respectively. In each time slot, each vessel selects an UAV or BS to offload a portion of task data. Each UAV and HAP offload part of the selected vessel’s data to HAP and satellite, respectively. Meanwhile, the BS, UAVs, HAP, and satellite compute buffered data from different vessels.

\subsection{Communication Model}

Following \cite{zwang2025double, sqi2025joint}, the vessel $v$-to-UAV $u$ (V2U) and vessel $v$-to-BS (V2B) links are modeled as the combination of path loss $h^L_{v,i}(n)=\frac{L_0}{d_{v,i}(n)^2}=\frac{L_0}{{H_i}^2+\| \boldsymbol{W_i}(n) - \boldsymbol{W_v}(n)\|^2}$ and Rician fading $h^R_{v,i}(n)=\sqrt{\frac{K_0}{1+K_0}}+\sqrt{\frac{1}{1+K_0}}o_{v,i}(n)$, $i \in \{u,b\}$, where $d_{v,i}(n)$ is the distance, $L_0$ is reference path loss, $o_{v,i}(n) \in \mathcal{CN}(0,1)$, and $K_0$ is Rician factor. The channel gain is given by $h_{v,i}(n)=h_{v,i}^{L}(n)|h_{v,i}^{R}(n)|^{2}$. For the link from UAV $u$-to-HAP (U2H), we denote the total path loss defined in \cite{wwu2025multi} by $L_{u,h}(n)$. The channel gain from HAP-to-satellite (H2S) is modeled as $h_{h,s}(n)=K_0 10^{-\frac{\eta_{h,s}(n)}{10}} G_h G_s$, where $\eta_{h,s}(n)$ signifies H2S large-scale fading \cite{wwu2025multi}, $G_h$ and $G_s$ are their respective antenna gains. Employing orthogonal frequency division multiple access (OFDMA) \cite{mdai2023latency, wwu2025multi}, the data rates of the V2U/B, U2H, and H2S links are respectively expressed as $R_{v,i}(n) = B_{v,i}(n) \log_2 \left(1+\frac{P_v(n)h_{v,i}(n)}{N_0 B_{v,i}(n)}\right)$, $R_{u,h}(n) = B_h \log_2 \left(1+\frac{P_u(n) G_u G_h 10^{-L_{u,h}(n)/10}}{ N_0 B_h}\right)$, and $R_{h,s}(n) = B_s \log_2\left(1+\frac{P_h(n) h_{h,s}(n)}{N_0 B_s}\right)$. $P_v(n)$, $P_u(n)$, and $P_h(n)$ are the transmit power of vessel $v$, UAV $u$, and HAP. $B_{v,i}(n)$, $B_h$, $B_s$ denote the bandwidth of each V2U/B, U2H, and H2S link. $G_u$ is UAV $u$'s antenna gain. With ultra-high-rate laser communications, transmission delay of the inter-satellite links (ISLs) is negligible and propagation delay is assumed as $T_p$ \cite{resswein2026constellation}. We focus on adaptive backlog control at handover and leave the study
 on ISL to future works.

\subsection{Computation and Task Queue Models}

At time slot $n$, the available computing resource (i.e., maximum processable CPU cycles) of server $i$ is $F_i(n), i \in \mathcal{I} = \{u,b,h,s\}$. The task data backlog on vessel $v$ is given by $Q_{v,0}(n) = Q_{v,0}(n-1) - \sum_{u \in \mathcal{U}} a_{v,u}(n-1) D_{v,u}^{tx}(n-1) - a_{v,b}(n-1) D_{v,b}^{tx}(n-1)$, where $D_{v,u}^{tx}$ and $D_{v,b}^{tx}$ are data amount offloaded to UAV $u$ and BS, $a_{v,u}$ and $a_{v,b}$ are binary indicators of vessel $v$'s offloading decision. The backlog of vessel $v$'s data on server $i$ is $Q_{v,i}(n) = Q_{v,i}(n-1) + D_{v,i}^{tx}(n-1) - D_{v,i}^{comp}(n-1) - \boldsymbol{1}_{\{ i \in \{u,h\} \}} \cdot D_{v,i,i'}^{tx}(n-1)$, $i' \in \{h,s\}$, where $D_{v,i}^{tx}$ denotes the total data volume of vessel $v$ received by server $i$, $D_{v,i}^{comp}$ is the data volume of vessel $v$ computed by server $i$, and $D_{v,i,i'}^{tx}(n)=$ (1) $a_{v,u,h}(n) D_{v,u,h}^{tx}(n), (i,i')=(u,h),$ (2) $a_{v,h,s}(n) D_{v,h,s}^{tx}(n), \hspace{0.1cm} (i,i')=(h,s)$, signifying the data of vessel $v$ offloaded from UAV $u$ to HAP or HAP to satellite. $a_{v,u,h} = 1$ indicates UAV $u$ chooses to offload vessel $v$'s data to HAP and $a_{v,h,s} = 1$ means HAP offloads vessel $v$'s data to satellite. The actual amount of transmitted data is bounded by link capacity and data backlog at each device.

\subsection{Problem Formulation}

We represent the total execution delay of vessel $v$'s task as $T_v = (N_v + 1) \cdot \tau$, where $N_v$ satisfies $Q_{v,0}(N_v) + \sum_{i \in \mathcal{I}} Q_{v,i}(N_v) > 0$ and $Q_{v,0}(N_v + 1) + \sum_{i\in \mathcal{I}} Q_{v,i}(N_v + 1) = 0$ simultaneously, indicating that its task data is just cleared from all queues at time slot $N_v + 1$. Our objective is to minimize the total task execution delay of all vessels by jointly optimizing the multi-layer task offloading decisions $\boldsymbol{a}$, UAV-BS bandwidth allocation $\boldsymbol{B}$, computing resource allocation $\boldsymbol{D^{comp}}$, and UAV trajectory $\boldsymbol{W_U}$. The overall problem can be formulated as

$\mathcal{P}_0$: \textbf{Overall Problem}
\begin{align*}
\min _{\substack {\boldsymbol{a}, \boldsymbol{B}, \\ \boldsymbol{D^{comp}}, \boldsymbol{W_U}}}
& \sum_{v \in \mathcal{V}} T_v\\
\text {s.t. } \hspace{0.33cm} &a_{v,u}(n), a_{v,b}(n), a_{v,u,h}(n), a_{v,h,s}(n) \in \{0,1\}, \\
&\forall v,u,n, \tag{1a}\\
& \sum_{u \in \mathcal{U}} a_{v,u}(n) + a_{v,b}(n) = 1, \forall v,n, \tag{1b}\\
& \sum_{v \in \mathcal{V}} a_{v,i}(n) B_{v,i}(n) \leq B_i(n), \forall i \in \{u,b\}, n, \hspace{-0.2cm}\tag{1c}\\
& B_{v,i}(n) \geq 0, \forall i \in \{u,b\},v,n, \tag{1d}\\
& \sum_{v \in \mathcal{V}} D_{v,i}^{comp}(n) C_{v} \leq F_i(n), \forall i,n, \tag{1e}\\
& D_{v,i}^{comp}(n) \leq Q_{v,i}(n) + D_{v,i}^{tx}(n), \\ 
& \forall v, i \in \{b,s\}, n, \tag{1f} \\
& D_{v,i}^{comp}(n) \leq Q_{v,i}(n) + D_{v,i}^{tx}(n) - D_{v,i,i'}^{tx}(n), \\
& \forall v, (i, i') \in \{(u, h), (h,s)\}, n, \tag{1g} \\
& D_{v,i}^{comp}(n)\geq0,\forall v,i,n, \tag{1h} \\
& \Delta d_u(n)/\tau \leq S_{u,max},\forall u, n,  \tag{1i} \\
& \| \boldsymbol{W_u}(n)-\boldsymbol{W_{u'}}(n)\|\geq d_{safe}, \forall u,u' \in \mathcal{U}, \\
& u\neq u', \forall n. \tag{1j}
\end{align*}

Regarding the constraints, (1a) and (1b) ensure the binary indicators and each vessel offloads data to one UAV or BS in each slot, respectively. (1c) and (1d) guarantee the bandwidth allocated to vessels is non-negative and bounded by total available spectrum of each UAV and BS. (1e) ensures the total allocated computing resources of each server do not exceed its available capacity. (1f)-(1h) guarantee the computed data size is bounded by available backlog and non-negative. (1i) enforces that the motion of UAV $u$ is bounded by its maximum speed $S_{u,max}$. (1j) is UAV collision-avoidance constraint, where $d_{safe}$ denotes minimum safe distance among UAVs. To tackle this mixed-integer nonlinear programming (MINLP) problem in dynamic SAGSIN, we first develop an efficient multi-layer task offloading scheme adaptive to real-time system conditions, and then iteratively optimize multidimensional system resources via block coordinate descent (BCD) framework.


\section{Satellite Handover-Aware Dynamic Task Offloading Algorithm for Multi-Layer SAGSIN}

Task offloading in SAGSIN involves twofold challenges: the abrupt shift in satellite resources caused by handover and the high complexity of task scheduling in dynamic multi-layer networks. To address these issues, this section first introduces the anticipatory strategy for satellite handover and then elaborates on the dynamic task offloading algorithm for SAGSIN.

\subsection{Anticipatory Satellite Handover Strategy}

The satellite offloading strategy based solely on the current satellite state can be myopic. If incoming satellite has significantly lower computing capacity, aggressive offloading before handover will lead to congestion afterwards. Conversely, if it offers stronger computing power, such strategies may conservative. An adaptive strategy accounting for both satellites is essential to smooth such abrupt changes.

\begin{figure}[!t]
\centering
\includegraphics[width=3.55 in]{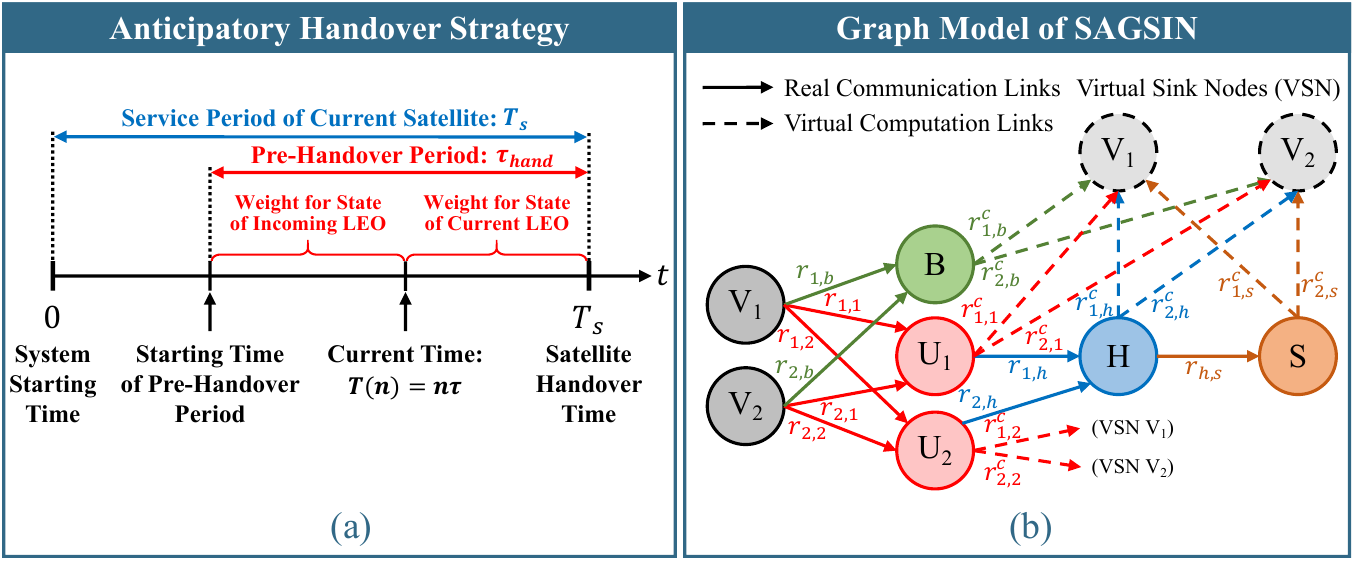}
\vspace{-12pt}
\caption{Satellite handover-aware multi-layer task offloading scheme: (a) Illustration of real-time joint satellite state for anticipatory handover strategy. (b) An example of graph model containing 2 vessel users, 2 UAVs, 1 BS, 1 HAP, and 1 LEO satellite, represented by node V, U, B, H, and S, respectively.}
\vspace{-15pt}
\label{offloading_fig}
\end{figure}

To this end, we design the real-time joint states of current and incoming satellites to guide offloading decision-making, adaptively regulating the data offloaded to satellites in advance of handover. A pre-handover period $\tau_{hand}$ is set prior to handover time $T_s$, as illustrated in Fig.~\ref{offloading_fig} (a). Once system time $T(n)$ enters this period, the joint states of available computing resources and backlog for satellites are respectively defined as
\begin{equation*}
\widetilde{F}_s(n) = \frac{T_s - T(n)}{\tau_{hand}} F_s(n) + \left(1-\frac{T_s - T(n)}{\tau_{hand}}\right) F_{s'}(n),  \tag{2a}
\end{equation*}
\begin{equation*}
\widetilde{Q}_{v,s}(n) \hspace{-1mm}=\hspace{-1mm} \frac{T_s - T(n)}{\tau_{hand}} Q_{v,s}(n) \hspace{-0mm}+\hspace{-0.5mm} \left(\hspace{-1mm} 1 \hspace{-1mm}-\hspace{-1mm}\frac{T_s - T(n)}{\tau_{hand}} \hspace{-0.5mm} \right) \hspace{-0.5mm} Q_{v,s'}(n), \tag{2b}
\end{equation*}
where $F_{s'}(n)$ and $Q_{v,s'}(n)$ are available computing resources and vessel $v$'s backlog at time slot $n$ on the incoming satellite. They are dynamic weighted sums of states for current and incoming satellites, where the weight of incoming one increases as handover approaches. These real-time joint states are employed to guide the task offloading strategy, thereby smoothing the transition and enhancing satellite resource utilization.

\subsection{Dynamic Task Scheduling in Multi-Layer SAGSIN}

The multi-layer architecture of SAGSIN leads to prohibitive complexity in optimizing task offloading decisions. To address this challenge, we leverage backpressure (BP) routing theory to decompose the combinatorial problem into elegant layer-wise task scheduling, thereby reducing complexity while adapting to multidimensional system dynamics.

To align with routing theory, the network is remodeled into the graph shown in Fig.~\ref{offloading_fig} (b). A virtual sink node (VSN) is established for each vessel $v$. We first homogenize the communication and computing capabilities, laying the foundation for designing the informative task scheduling indicators. We define the rate of each real communication link (i.e., $r_{v,u}$, $r_{v,b}$, $r_{u,h}$, and $r_{h,s}$) as the corresponding channel capacity $R$ times $\tau$, and the rate of virtual computation link between server $i$ and VSN $v$ as $r_{v,i}^c = F_i / C_v$. In this way, the rate $r_e$ of each link $e$ is unified as the number of transmitable/computable bits per time slot. The distance of each link $e$ is $w_e = \overline{r} \cdot r_{max} / r_e$, where $\overline{r}$ and $r_{max}$ are average and maximum link rates in the system.

To guide data towards servers with low congestion and abundant resources, we define the real-time Pressure Index (PI) of vessel $v$ and server $i$ regarding vessel $v$'s data as
\begin{equation*}
\left\{
\begin{aligned}
J_{v,0}(n) &= Q_{v,0}(n) + w_{v,0}^{min}(n), \\
J_{v,i}(n) &= Q_{v,i}(n) + w_{v,i}^{min}(n), \\
\end{aligned}
\right. \tag{3}
\end{equation*}
where $w_{v,0}^{min}(n)$ and $w_{v,i}^{min}(n)$ are the shortest distances from that node to VSN $v$. The PI serves as an informative indicator that integrates task congestion level, computing capacity, rate of links to high-capacity servers, and network topology. As the driving force for offloading, the Pressure Differential (PD) of vessel $v$'s data on links between vessel $v$ and UAV $u$/BS, UAV $u$ and HAP, HAP and satellite are respectively given by
\begin{equation*}
\left\{
\begin{aligned}
& \Delta J_{v \to i}(n) = J_{v,0}(n) - J_{v,i}(n), \forall i \in \{u,b\}, \\
& \Delta J_{v,u \to h}(n) = \max\{J_{v,u}(n) - J_{v,h}(n), 0\}, \\
& \Delta J_{v,h \to s}(n) = \max\{J_{v,h}(n) - J_{v,s}(n), 0 \}.
\end{aligned}
\right. \tag{4}
\end{equation*}
Then, the offloading decision of each vessel $v$ is given by
\begin{equation*}
\boldsymbol{a_{v,i}}(n)^* = \underset{ i \in \{u, b\}} {\operatorname{\operatorname{argmax}}} \boldsymbol{a_{v,i}} (n)^T \left[\Delta\boldsymbol{J_{v \to i}}(n) \odot \boldsymbol{r_{v,i}}(n) \right], \tag{5}
\end{equation*}
where $\boldsymbol{a_{v,i}}(n) = [a_{v,1}(n), \dots, a_{v,U}(n), a_{v,b}(n)]^T$ subject to (1b), $\Delta\boldsymbol{J_{v \to i}}(n) = [ \Delta J_{v \to 1}(n), \dots, \Delta J_{v \to U}(n), \Delta J_{v \to b}(n) ]^T$, and $\boldsymbol{r_{v,i}}(n) = [ r_{v,1}(n), \dots, r_{v,U}(n), r_{v,b}(n) ]^T$. This mechanism leads each vessel to select the UAV or BS offering the best combination of low load and high throughput, thereby avoiding server overload while fully exploiting high-rate channels. The offloading decision of each UAV $u$ and HAP is expressed as 
\begin{equation*}
\boldsymbol{a_{v,i,i'}}(n)^* = \underset{ v \in \mathcal{V}} {\operatorname{\operatorname{argmax}}}\boldsymbol{a_{v,i,i'}}(n)^T \Delta\boldsymbol{J_{v, i \to i'}}(n), \tag{6}
\end{equation*}
where $\Delta\boldsymbol{J_{v, i \to i'}}(n) = [ \Delta J_{1, i \to i'}(n), \dots, \Delta J_{V, i \to i'}(n) ]^T$ and $\boldsymbol{a_{v,i,i'}}(n) = [ a_{1,i,i'}(n), \dots, a_{V,i,i'}(n) ]^T$ subject to $\sum_{v \in \mathcal{V}} a_{v,i,i'}(n) \leq 1$, $(i,i') \in \{(u,h), (h,s)\}$. Formula (6) indicates that each UAV or HAP selects a vessel with the most urgent offloading demand (e.g., severe backlog or the possibility of quickly reaching upper-layer server with more resources) and transmits its data to HAP or satellite.

By shifting from global coordination to layer-wise pressure-driven forwarding, our scheme reduces the complexity from exponential level to approximate $\mathcal{O}(VU)$. Furthermore, since the PI is updated at each time slot based on instantaneous system states, it possesses strong adaptability to the high dynamics of SAGSIN environments.


\section{Joint UAV-BS Bandwidth, UAV Trajectory, and Computing Resource Optimization}

Building upon the layer-wise task scheduling strategy derived from a macroscopic perspective of network topology and system states, this section focuses on the fine-grained orchestration of underlying resources to further enhance the system performance. Specifically, we jointly optimize the inter-coupled UAV-BS bandwidth allocation, UAV trajectories, and computing resource allocation via a BCD framework.

\subsection{Bandwidth Allocation Policy of UAVs and BS}

This subproblem aims to maximize the total transmission rate of communication links between each UAV/BS and the vessels it serves by optimizing bandwidth allocation, thereby boosting the volume of offloaded vessel data and reducing overall task delay. This subproblem can be formulated as

$\mathcal{P}_1$: \textbf{UAV and BS Bandwidth Allocation}
\begin{align*}
\max _{\substack {B_{v,i}(n)}}
& \sum_{i \in \{u,b\}} \sum_{v \in \mathcal{V}} a_{v,i}(n) B_{v,i}(n) \log_2 \left( 1 + \frac{P_v(n) h_{v,i}(n) } {B_{v,i}(n) N_0} \right) \\
\text {s.t. } \hspace{0.1cm} & \hspace{0.3cm} \text{(1c) and (1d)},
\end{align*}
With determined vessel offloading decisions, the objective function and constraints are strictly concave w.r.t. $B_{v,i}(n)$. Thus, $\mathcal{P}_1$ is a convex optimization problem. The optimal bandwidth allocation of BS and UAVs are expressed as 
\begin{equation*}
B_{v,i}^*(n) \hspace{-0.5mm}=\hspace{-0.5mm} \frac{ P_v(n) h_{v,i}(n)} {\sum_{v \in \mathcal{V}_i(n)} P_v(n) h_{v,i}(n)} B_i(n), v \in \mathcal{V}_i(n), i \in \{b,u\}, \tag{7}
\end{equation*}
where $\mathcal{V}_i(n)$ signifies set of vessels associated with server $i$. Proofs and derivations are omitted due to space limitation.


\subsection{UAV Trajectory Optimization}

Beyond resource dimension, the mobility of UAVs provides a crucial degree of freedom to dynamically reconstruct the network topology. By optimizing their trajectories, we directly modulate the vessel-to-UAV channel states. Specifically, the trajectory planning aims to strategically position UAVs to maximize the aggregate data rate and throughput for associated vessels while strictly adhering to kinematic and safety constraints. Accordingly, this subproblem is formulated as

$\mathcal{P}_2$: \textbf{UAV Trajectory Optimization}
\begin{align*}
\max _{\substack {\boldsymbol{W_u}(n)}}
& \sum_{u \in \mathcal{U}} \sum_{v \in \mathcal{V}} a_{v,u}(n) B_{v,u}(n) \log_2 \left( 1 + \frac{P_v(n) h_{v,u}(n) } {B_{v,u}(n) N_0} \right)\\
\text {s.t. } & \text{(1i) and (1j)}.
\end{align*}

The non-convexity of the objective function and limit (1j) poses the primary challenge. To tackle this, we employ the successive convex approximation (SCA) technique. By applying the first-order Taylor expansion of data rate $R_{v,u}(n)$ at the given point $\phi_{v,u}^{(k)}(n) = \| \boldsymbol{W_u^{(k)}}(n) - \boldsymbol{W_v}(n) \|^2$ in the $k$-th iteration, we can obtain its global lower-bound as
\begin{equation*}
\widetilde{R}_{v,u}(n)=R_{v,u}^{(k)}(n)+\nabla R_{v,u}^{(k)}(n) (\phi_{v,u}(n)-\phi_{v,u}^{(k)}(n)), \tag{8}
\end{equation*}
where $R_{v,u}^{(k)}(n)$ and $\nabla R_{v,u}^{(k)}(n)$ are the data rate and first-order derivative of $R_{v,u}(n)$ w.r.t. $\phi_{v,u} (n)$ at $k$-th iteration, respectively. Similarly, the square of inter-UAV distance $\| \boldsymbol{W_u}(n)-\boldsymbol{W_{u'}}(n)\|^2 $ is globally lower bounded by first-order Taylor expansion at given point $ \psi_{u,u'}^{(k)}= \boldsymbol{W_u^{(k)}}(n)-\boldsymbol{W_{u'}^{(k)}}(n)$ as 
\begin{equation*}
{\widetilde{d}_{u,u'}(n)}^2 \hspace{-0.5mm}=\hspace{-0.5mm} \|\psi_{u,u'}^{(k)}\|^{2} \hspace{-0.5mm}+\hspace{-0.5mm} 2 \psi_{u,u'}^{(k) \hspace{1.5mm} T} ((\boldsymbol{W_u}(n)-\boldsymbol{W_{u'}}(n))-\psi_{u,u'}^{(k)}). \tag{9}
\end{equation*}
Then, we convert $\mathcal{P}_2$ into a series of approximated convex subproblems expressed as

$\mathcal{P}_2'$: \textbf{Convex Approximation Problem of $\mathcal{P}_2$}
\begin{align*}
\max _{\substack {\boldsymbol{W_u}(n)}}
& \sum_{u \in \mathcal{U}} \sum_{v \in \mathcal{V}_u(n)} \widetilde{R}_{v,u}(n)\\
\text {s.t. } & \text{(1i)}, \\
& {\widetilde{d}_{u,u'}(n)}^2 \geq {d_{safe}}^2, \forall u,u' \in \mathcal{U}, u\neq u', \tag{10}
\end{align*}
which can be solved by CVX solver iteratively until converging to a near-optimal solution. This ensures that the positions of UAVs evolve synergistically with vessel mobility and dynamic task demands, effectively bridging the gap between topology optimization and physical-layer connectivity.


\subsection{Computing Resource Allocation}

\begin{algorithm}[!t]
    \caption{Computing Resource Allocation Algorithm.}
    \label{alg:computing}
    \renewcommand{\algorithmicrequire}{\textbf{Input:}}
    \renewcommand{\algorithmicensure}{\textbf{Output:}}
    
    \begin{algorithmic}[1]

        \begin{small}
        \REQUIRE $\mathcal{V}$, $Q_{v,i}(n)$, $F_i(n)$, $r_{v,i}^c(n)$, $D_v$, and $C_v$.   
        \ENSURE Computing resource allocation strategy $\boldsymbol{F_{v,i}}(n) = \{F_{v,i}(n)|\forall v\}$ for server $i$.     

        \STATE Initialize the pending vessel set $\mathcal{V}_p = \mathcal{V}$, the assigned vessel set $\mathcal{V}_a = \emptyset$, and the remaining computing resources $F_i^{re}(n) = F_i(n)$.

        \WHILE{$\mathcal{V}_p \neq \emptyset $}
        
            \STATE Allocate computing resource to each vessel $v \in \mathcal{V}_p $ based on $F_{v,i}(n) = \frac{Q_{v,i}(n) / r_{v,i}^c(n)} {\sum_{v \in \mathcal{V}_p} Q_{v,i}(n) / r_{v,i}^c(n)}\cdot F_i^{re}(n)$.

            \IF{$F_{v,i}(n) \leq F_{v,i}^{max}(n), \forall v \in \mathcal{V}_p$}

                \STATE Obtain the final solution $\boldsymbol{F_{v,i}}(n) = \{F_{v,i}(n)|\forall v\}$.
                \STATE \textbf{break}

            \ELSE

                \FOR{$v \in \mathcal{V}_p$ such that $F_{v,i}(n) > F_{v,i}^{max}(n)$}
                    
                    \STATE Set its $F_{v,i}(n) \leftarrow F_{v,i}^{max}(n)$.
                    \STATE Move this vessel $v$ from $\mathcal{V}_p$ to $\mathcal{V}_a$.
    
                \ENDFOR

                \STATE Update $F_i^{re}(n) \leftarrow  F_i(n) - \sum_{v \in \mathcal{V}_a} F_{v,i}(n)$.

            \ENDIF

        \ENDWHILE

        \end{small}

    \end{algorithmic}
\end{algorithm}

As the final stage of task execution pipeline in SAGSIN, dynamic allocation of computing resources governs the terminal processing rate for each vessel's data. Denoting the computing resource that server $i$ allocates to vessel $v$ at time slot $n$ as $F_{v,i}(n)$ (CPU cycles), constraints (1e)-(1h) can be rewritten as 
\begin{equation*}
\sum_{v \in \mathcal{V}} F_{v,i}(n) \leq F_i(n) , \forall i,\tag{11a}
\end{equation*}
\begin{equation*}
0 \leq F_{v,i}(n) \leq F_{v,i}^{max}(n), \forall v,i,\tag{11b}
\end{equation*}
where we denote maximum required resources of vessel $v$ at server $i$ as $F_{v,i}^{max}(n) = C_v \cdot ( Q_{v,i}(n) + D_{v,i}^{tx}(n) - 
 \boldsymbol{1}_{\{ i \in \{u,h\} \}} \cdot D_{v,i,i'}^{tx}(n) ), \forall i $. Each server should allocate resources based on backlog of vessels in its buffer while meeting above limits. To achieve this, we design a demand-driven computing resource allocation method as detailed in \textbf{Algorithm~\ref{alg:computing}}. Only the key parameters of this subproblem are listed in the input for brevity. Computing resources are allocated in proportion to completion time of each vessel's data on this server, thereby prioritizing vessels with severe backlog while avoiding resource waste. After gaining the solution, $D_{v,i}^{comp}(n) = F_{v,i}^*(n) / C_v , \forall v,i$.

The steps of overall approach is outlined in \textbf{Algorithm~\ref{alg:overall}}. At the beginning, we initialize system parameters and update backlog queues of all vessels and servers. Then the multi-layer task offloading decisions are obtained through the method detailed in Section III. According to dependencies among subproblems, we iteratively optimize the UAV-BS bandwidth allocation, UAV trajectory planning, and computing resource allocation in sequence, until convergence or the maximum number of iterations is reached. The superscript $j$ denotes the solution at the start of the $j$-th iteration.

\begin{algorithm}[!t]
    \caption{Overall Dynamic Task and Resource Scheduling Method for Green SAGSIN.}
    \label{alg:overall}
    \renewcommand{\algorithmicrequire}{\textbf{Input:}}
    \renewcommand{\algorithmicensure}{\textbf{Output:}}

    \begin{algorithmic}[1]

        \begin{small}
    
        \REQUIRE The optimized UAV positions $\boldsymbol{W_u}(n-1)$ in previous time slot and all remaining parameters in current time slot.   
        \ENSURE Optimization results $\boldsymbol{a}(n)$, $\boldsymbol{B}(n)$, $\boldsymbol{D^{comp}}(n)$, and $\boldsymbol{W_U}(n)$.
        
        \STATE Update backlog of all devices or set $Q_{v,0}(0) = D_v$ and $Q_{v,i}(0) = 0$ at initial time slot, and initialize iteration number $j=0$.
        \STATE Obtain $\boldsymbol{a}(n)$ through the method detailed in Section III.

        \REPEAT
            \STATE Solve $\mathcal{P}_1$ to obtain $\boldsymbol{B}(n)^{j+1}$ with given $\boldsymbol{W_U}(n)^j$ and $\boldsymbol{D^{comp}}(n)^j$.
            
            \STATE Solve $\mathcal{P}_2'$ to obtain $\boldsymbol{W_U}(n)^{j+1}$ with given $\boldsymbol{B}(n)^{j+1}$ and $\boldsymbol{D^{comp}}(n)^j$.
            
            \STATE Execute \textbf{Algorithm~\ref{alg:computing}} to obtain $\boldsymbol{D^{comp}}(n)^{j+1}$ with given $\boldsymbol{B}(n)^{j+1}$ and $\boldsymbol{W_U}(n)^{j+1}$.
            
            \STATE Update the value of objective function.
            
            \STATE Update $j=j+1$.

        \UNTIL The objective value converges or $j>j_{max}$.

        \end{small}

    \end{algorithmic}
\end{algorithm}


\section{Simulation Results}

In this section, we conduct simulation experiments to evaluate the performance of the proposed method. We consider a 2 km$\times$2 km ocean area and uniformly deploy six UAVs with maximum speed of 15 m/s and minimum safe distance of 5 m \cite{sqi2025joint}. Vessels are randomly distributed, with the velocity ranging from 5 to 15 m/s \cite{wwu2025multi}. A coastal BS is deployed at the position of (0, 1km) and HAP maintains a quasi-stationary hover at center of the region \cite{wli2026efficient}. Similar to \cite{zwang2025double, wwu2025multi}, the key parameter setup is summarized in Table~\ref{tab:table1}. To validate the effectiveness of the proposed DASH method, we compare it with the following benchmark schemes:

\begin{itemize}
    \item \textbf{DASH w/o HO}: The proposed DASH method without anticipatory satellite handover strategy.
    \item \textbf{HACO}: A multi-HAP-assisted computation offloading algorithm for SAGSIN proposed in \cite{wwu2025multi}, targeting the minimization of overall task execution delay.
    \item \textbf{FLEC}: A four-layer edge computing scheme that aims to minimize system task costs proposed in \cite{xwang2024amtos} with a similar hierarchical architecture to this study.
\end{itemize}



\begin{table}[!h]
\caption{Simulation Parameters}
\label{tab:table1}
\centering
\begin{tabular}{|m{1.8cm}<{\centering}|m{3.2cm}<{\centering}|m{1.9cm}<{\centering}|}
\hline
\textbf{Parameter} & \textbf{Definition} & \textbf{Value}\\
\hline
$V$ & Number of vessels & 10--30 \\
\hline
$D_v$ & Data size of vessels & 2--10 Mb \\
\hline
$C_v$ & Computational density & 100--2000 \\
\hline
$N_0$ & Noise power spectral density & -174 dBm/Hz\\
\hline
$L_0$ & Reference path loss & -30 dB\\
\hline
$H_u, H_b, H_h$ & Height of UAV, BS, HAP & [0.1, 0.03, 20] km\\
\hline
$d_{h,s}$ & H2S distance & 784 km\\
\hline
$F_u, F_b, F_h,F_s$ & Computing resources of each server & [1, 3, 3, 10] ×$10^8$ cycles/slot \\
\hline
$B_u, B_b, B_h, B_s$ & Bandwidth of each server & [10, 20, 20, 50] MHz\\
\hline
$P_v, P_u, P_h$ & Transmit power & [1, 1.5, 2.5] W \\
\hline
$G_u, G_h, G_s$ & Antenna gains & [25, 30, 35] dBi \\
\hline
$\tau$ & Duration of time slot & 0.1 s\\
\hline

\end{tabular}
\end{table}

\begin{figure}[!t]
\centerline{\includegraphics[width=2.5 in]{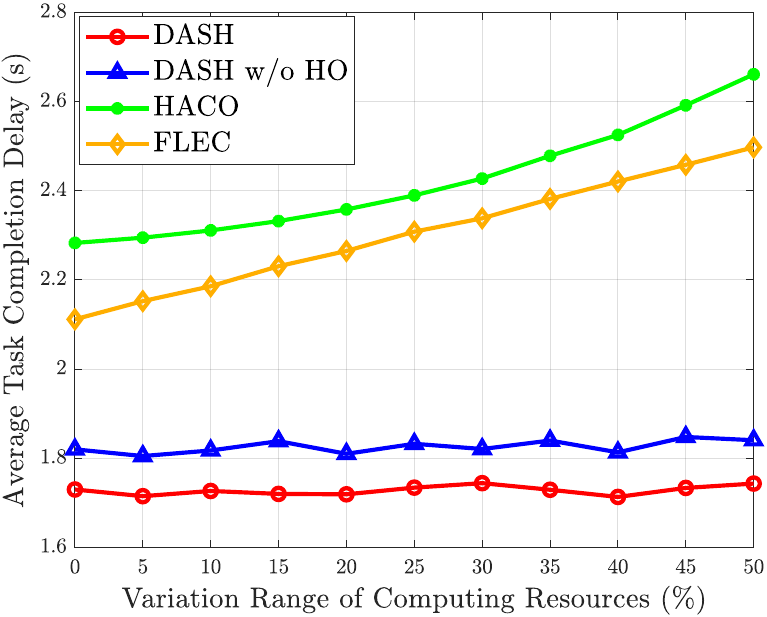}}
\caption{The comparison of average task completion delay among different methods with varying
fluctuation range of available computing resources.}
\vspace{-5mm}
\label{exp_comp_flu}
\end{figure}

Fig.~\ref{exp_comp_flu} illustrates the trend of average task completion delay for different methods with the percentage variation range of available computing resources for servers across time slots. DASH adaptively adjusts the fine-grained task and resource scheduling strategy based on real-time available computing resources of all servers as well as the incoming satellite, achieving the lowest average task completion delay. When the incoming satellite has limited computing capacity, DASH w/o HO cannot proactively constrain the offloaded task volume to satellite, leading to post-handover congestion and increased satellite computation delay. The one-shot task and resource scheduling of HACO and FLEC leads to inconsistency in the available computing resources of servers between the decision-making time and the time when tasks arrive at destination servers. Upon task arrival, a reduction in available computing capacity increases computation delay, whereas an increased resource cannot be exploited. Consequently, their average task delays grow monotonically with intensifying computing resource fluctuations. DASH adaptively routes task data based on instantaneous server states, which effectively absorbs the impact of resource fluctuations, reducing the average delay by around 28\% and 25\% compared to HACO and FLEC, respectively.

\begin{figure}[!t]
\centerline{\includegraphics[width=2.5 in]{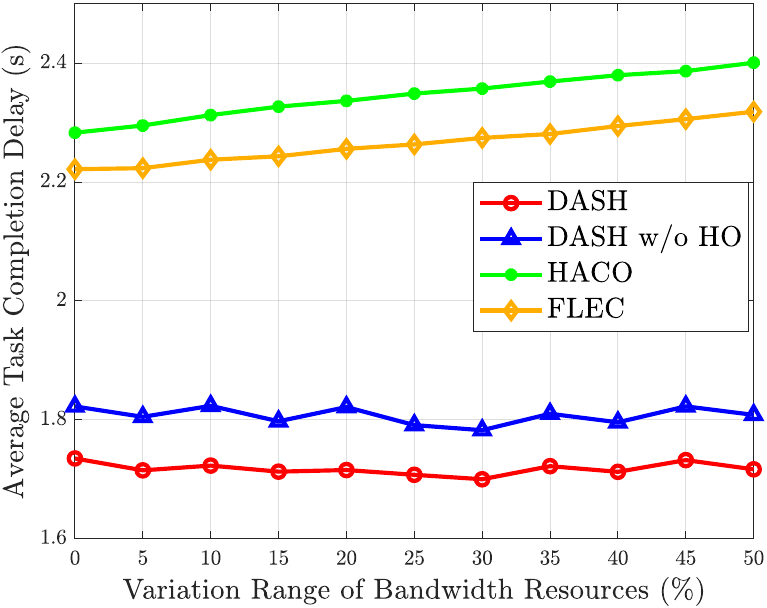}}
\caption{The comparison of average task completion delay among different methods with varying
fluctuation range of available bandwidth resources.}
\vspace{-3mm}
\label{exp_bw_flu}
\end{figure}

Fig.~\ref{exp_bw_flu} depicts the average task completion delay among different methods with respect to the percentage variation range of available bandwidth for wireless links in the system across time slots. For the proposed DASH method and its ablation variant DASH w/o HO, task offloading and bandwidth resource allocation are dynamically adjusted based on real-time link states, including available bandwidth, large and small-scale fading, and channel capacities, enabling strong adaptability to communication dynamics and achieving low and stable average task delay with bandwidth variation. DASH reduces the average task delay respectively by around 26\% and 23\% compared to HACO and FLEC, since their one-shot scheduling strategies fail to adaptively reroute task when channel conditions deteriorate or reallocate resources when bandwidth becomes abundant. Compared with more dominant computing resources, which govern the terminal processing rates, the ‌intensification‌ of bandwidth fluctuations has a relatively smaller impact on the increase of their average task delay.

\begin{figure}[!t]
\centerline{\includegraphics[width=2.5 in]{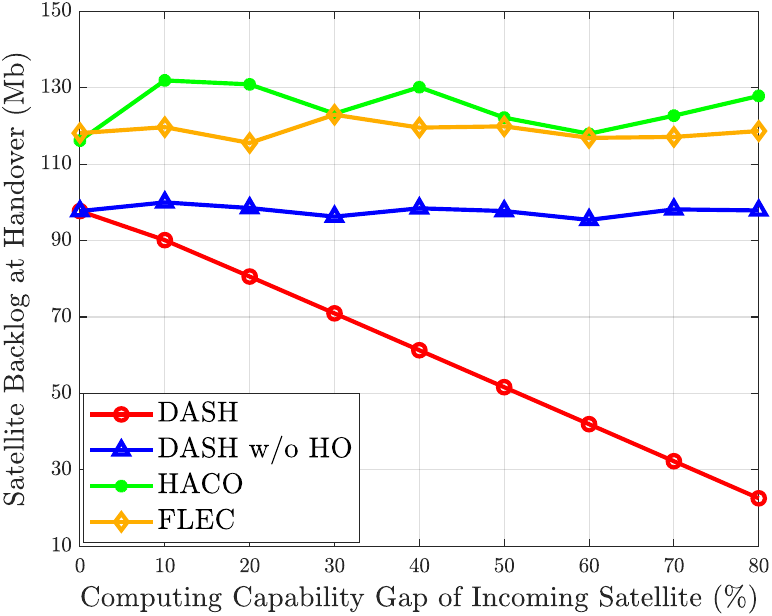}}
\caption{The comparison of satellite backlogged data at handover among different methods with varying
computing resource gap of incoming satellite.}
\vspace{-6mm}
\label{exp_backlog}
\end{figure}

Fig.~\ref{exp_backlog} shows the amount of backlogged task data on the current satellite at handover (i.e., data to be transferred to incoming satellite) versus the deficit percentage in the average available computing resource of the incoming satellite relative to the current one. Due to the anticipatory handover strategy, DASH markedly reduces the amount of backlogged data at handover as the incoming satellite’s available computing capacity continuously decreases below that of the current satellite, thereby effectively alleviating post-handover congestion. Benefiting from fine-grained and more balanced task and resource scheduling, DASH w/o HO achieves lower satellite backlog at handover than HACO and FLEC. However, since all those three methods determine strategies solely based on the current satellite state, they lack adaptability to changes in the computing capability of incoming satellite. Consequently, their handover backlogs remain insensitive to the increasing gaps in the next satellite’s computing resources.


\section{Conclusion}

This paper proposed a dynamic task and resource scheduling method for green SAGSIN to minimize task execution delay for vessel users. To address the complexity and high dynamics of multi‑layer task offloading, we developed a layer‑wise offloading scheme that adapts to real‑time system states, along with an anticipatory satellite handover policy that prevents post‑handover congestion. Additionally, we jointly optimized UAV‑BS bandwidth allocation, UAV trajectories, and computing resources to accelerate task completion and enhance resource utilization. Experimental results show that the proposed method significantly reduces task delay under fluctuating system resources and adaptively controls satellite backlog during handover compared with benchmarks.


\end{document}